\documentclass[epj]{webofc}
\usepackage[varg]{txfonts}   % Web of Conferences font
\usepackage{graphicx,color} % for figures
\usepackage{bm}       % bold math 
\usepackage{amsmath}  
\usepackage{amssymb}
\woctitle{21st International Conference on Few-Body Problems in Physics}
\begin{document}
\title{Advances in the
{\it ab initio} 
description of nuclear three-cluster systems}
\author{Carolina Romero-Redondo\inst{1}\fnsep\thanks{\email{romeroredond1@llnl.gov}} \and
        Sofia Quaglioni\inst{1}\fnsep\thanks{\email{quaglioni1@llnl.gov}} \and 
        Petr Navr\'atil\inst{2}\fnsep\thanks{\email{navratil@triumf.ca}}\and
	Guillaume Hupin\inst{3}\fnsep\thanks{\email{hupin@ipnorsay.in2p3.fr}} 
        % etc.
}

\institute{Lawrence Livermore National Laboratory, P.O. Box 808, L-414, Livermore, California 94551, USA
\and
           TRIUMF, 4004 Wesbrook Mall, Vancouver, British Columbia, V6T 2A3, Canada
\and
      Institut de
Physique Nucl\'eaire, Universit\'e Paris-Sud, IN2P3/CNRS, F-91406 Orsay
Cedex, France           
          }

\abstract{
We introduce the extension of the $ab~initio$
 no-core shell model with continuum to 
describe three-body
cluster systems.  We present results for the ground state of
$^6$He and show improvements with respect to the description obtained within the no-core shell model 
and the no-core shell model/resonating group methods.  
}
\maketitle
\section{Introduction}
\label{intro}
The $ab~initio$ no-core shell model/resonating group method (NCSM/RGM) was
presented in \cite{QuaPRL08,QuaPRC09} as a technique that is able to
describe both structure and reactions in light nuclear systems. Within this approach, the wave function
is expanded in a continuum cluster basis using the resonating group method with realistic 
interactions and a consistent $ab~initio$ description of the nucleon clusters.

The method was first introduced in detail for two-body cluster bases and has
been shown to work efficiently in different systems \cite{QuaPRL08,QuaPRC09,NaVPRC11,NavPRL12}. Later,
the expansion of the method 
for three-cluster systems was introduced in \cite{Quaglioni:2013kma,Romero-Redondo:2014fya}. 

%The extension of the NCSM/RGM approach 
The capability of {\it ab initio} methods 
to properly describe three-body cluster
states is essential for the study of nuclear systems that present
such configuration. This type of systems appear, $e.g$,  in
structure problems of
two-nucleon halo nuclei such as $^6$He and $^{11}$Li, resonant systems such as $^5$H or
transfer reactions with
three fragments in their final states such as $^3$H($^3$H,2n)$^4$He or
$^3$He($^3$He,2p)$^4$He.

Despite the success of the NCSM/RGM in describing the long range behavior of the wave functions,
it has been shown that it has limitations when it comes to accurately account for short range 
correlations, which is   
necessary to achieve a complete description of the system. This is due to fact that to account for   
such correlations, several excited states of the nuclear clusters must be included in the basis, 
resulting in an increase of the problem size that goes
beyond current computational capabilities. This limitation has been overcome by introducing the
{\it ab initio} no-core shell model with continuum (NCSMC). With this method, the wave function is
written as a superposition of both continuum NCSM/RGM cluster states and discrete eigenstates of the
compound system obtained
with the no-core shell model (NCSM). The latter eigenstates compensate for the
missing cluster excitations improving the description of short range correlations.  
 
The NCSMC was first introduced in \cite{PhysRevLett.110.022505,PhysRevC.87.034326} for binary 
systems. Its expansion to three-cluster systems was recently achieved and we show here 
the first results for the $^6$He ground state (g.s).

\section{Formalism} 
\label{sec-1}

In the NCSMC, the ansatz for the three-cluster many-body wave function is given by
 
\begin{equation}
\label{eq:trialwf}
        |\Psi^{J^\pi T}\rangle  =\sum_{\lambda}c_{\lambda}|A\lambda J^{\pi}T\rangle + \sum_{\nu} \iint dx \, dy \,   
x^2\, y^2 \, G_{\nu}^{J^\pi T}(x,y) \, \hat {\mathcal A}_\nu\, |\Phi^{J^\pi T}_{\nu x y} \rangle \nonumber \,,  
        % Here I use G rather than \chi because these are not yet the (orthonormal) Schroedinger wave functions.
\end{equation}
%where $c_{\lambda}$ are the discrete coefficients of the expansion and $G_{\nu}^{J^\pi T}(x,y)$ 
where $c_{\lambda}$ and $G_{\nu}^{J^\pi T}(x,y)$ are, respectively, discrete and  
 continuous variational amplitudes,  $|A\lambda J^{\pi}T\rangle$ are the NCSM eigenstates 
labeled by the set of quantum number $\lambda$,   %of the integration variables $x$ and $y$, 
$\hat {\mathcal A}_\nu$ is an appropriate intercluster antisymmetrizer introduced to exactly preserve 
the Pauli exclusion principle, and

\begin{align}
         |\Phi^{J^\pi T}_{\nu x y} \rangle  = & 
        \Big[\Big(|A-a_{23}~\alpha_1I_1^{\pi_1}T_1\rangle 
        \left (|a_2\, \alpha_2 I_2^{\pi_2} T_2\rangle |a_3\, \alpha_3 I_3^{\pi_3}T_3\rangle \right)^{(s_{23}T_{23})}\Big)^{(ST)} 
        \left(Y_{\ell_x}(\hat{\eta}_{23})Y_{\ell_y}(\hat{\eta}_{1,23})\right)^{(L)}\Big]^{(J^{\pi}T)} \nonumber \\
        & \times \frac{\delta(x-\eta_{23})}{x\eta_{23}} \frac{\delta(y-\eta_{1,23})}{y\eta_{1,23}}\,,
        \label{eq:3bchannel}    
\end{align}
are three-body cluster channels of total angular momentum $J$, parity $\pi$ and isospin $T$ where
 $\nu$ represents a set of quantum numbers that 
describes the channel within the cluster basis.  
Here, $|A-a_{23}~\alpha_1I_1^{\pi_1}T_1\rangle$, $|a_2\, \alpha_2 I_2^{\pi_2} T_2\rangle$ 
and $|a_3\, \alpha_3 I_3^{\pi_3} T_3\rangle$ denote the microscopic (antisymmetric) 
wave functions of the three nuclear fragments 
calculated within the NCSM. The Jacobi coordinates describing the relative positions of the clusters are
denoted by $\eta_{23}$ and $\eta_{1,23}$.

We calculate the unknowns of the NCSMC wave function [$c_{\lambda}$ and $G_{\nu}^{J^\pi T}(x,y)$]  
by solving the orthogonalized coupled equations obtained by projecting the
Schr\"odinger equation on the model space spanned by NCSM eigenstates and the NCSM/RGM basis. 
% (see \cite{Quaglioni:2013kma}) and
%coupling terms arsing from the coupling of both bases. 
Those equations are solved by means of the 
microscopic R-matrix method in a Lagrange mesh \cite{Hesse199837}. Details on the procedure will be available in 
\cite{NCSMC3B}.  
 
%\subsection{Subsection title}
\label{sec-2}

%
%
%For figure with sidecaption legend use the `sidecaption' environment as
%given in Fig.~\ref{fig-2}
%\begin{figure}
%\centering
%\sidecaption
%\includegraphics[width=5cm,clip]{name_fig2.eps}
%\caption{Please write your figure caption here. It will appear adjacent to
%the figure.}
%\label{fig-2}       % Give a unique label
%\end{figure}

%\section{Use of tables}
%
%

\section{Application to $^6$He}

\begin{table}
\centering
\caption{Energy (in MeV) for the $^6$He ground state using the NCSM/RGM, NCSM and NCSMC approaches at $N_{max}$=12. For the NCSM we also show the extrapolated 
value to $N_{max}\to \infty$.}
\label{tab-1}       % Give a unique label
% For LaTeX tables you can use
\begin{tabular}{cccc}
\hline
$N_{\rm max}$  & NCSM/RGM & NCSM & NCSMC\\
\hline
%6&  $-28.907$ & $-27.705$ \\
8&  $-28.62$ & $ -28.95$ & $-29.69$\\ 
10&  $-28.72$ & $-29.45$& $-29.86$ \\
12& $-28.70$ & $-29.66$ & $-29.86$ \\
\hline
Extrapolation & --- & $-29.84(4)$ & --- \\
%   Approach
%  &
%%  &  E$_{g.s}$($^4$He)
%  &  E$_{g.s}$($^6$He)  \\\hline
%NCSM/RGM & ($N_{\rm max}$=12)       & $-28.70$ MeV \\
%NCSM & ($N_{\rm max}$=12) &  $-29.75$ MeV\\
%NCSM & (extrapolated) &  $-29.84(4)$ MeV\\
%NCSMC & ($N_{\rm max}$=12)   & $-29.86$ MeV \\
\hline
\end{tabular}
% Or use
%\vspace*{5cm}  % with the correct table height
\end{table}

The lightest Borromean nucleus is $^6$He ~\cite{Tanihata:1995yv, PhysRevLett.55.2676}, 
 formed by an $^4$He core and two halo neutrons. It is, therefore,
 an ideal first candidate to be studied within a three-body formalism. Hence,
it was used as a test case when the NCSM/RGM formalism for three-cluster 
dynamics was introduced in \cite{Quaglioni:2013kma,Romero-Redondo:2014fya} and here
is studied again in order to perform a benchmark with such results. 
In this first calculation, we describe the $^4$He core only by its g.s.
 wave function and couple the three-cluster basis
with the $^6$He g.s. eigenstate obtained through the NCSM.

We used the same potential as in \cite{Quaglioni:2013kma,Romero-Redondo:2014fya}, i.e.,
 the similarity-renormalization-group
 (SRG) \cite{SRG,SRG2} evolved potential obtained from the
 chiral N$^3$LO NN interaction \cite{NN}
with $\Lambda_{\mbox{\small{SRG}}}$ = 1.5 fm$^{-1}$. With this soft potential the binding
energy can be
accurately computed by extrapolating the NCSM results to $N_{max}\to \infty$, hence providing 
a good benchmark for the newly implemented NCSMC. 

% for The set of equations (\ref{RGMrho}) are solved for different
%channels using both bound and continuum asymptotic conditions. 
%We find only one bound state, which appears  in the
%$J^{\pi}T=0^+1$ channel and corresponds to the $^6$He ground state. 

\begin{figure}
\centering
\includegraphics[width=5.5cm,clip,angle=-90]{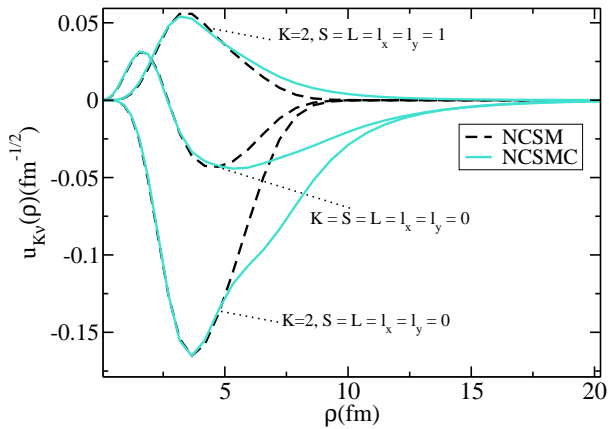}
\caption{Most relevant hyperradial contributions to the $^6$He g.s. wave function.  
Both the contribution from the 
NCSM wave function and the total NCSMC wave function are shown for a $N_{max}=6$ model space. The figure 
shows how the addition of the three-cluster basis within the NCSMC compensates the limitations of the
NCSM to obtain an extended wave function characteristic of two-neutron halo nuclei.
The hyperradial wave functions $u_{K\nu}(\rho)$ are the coefficients of the wave function when expanded
in the hyperspherical basis, where $K$ represents the hypermomentum.
}  
\label{fig-1}       % Give a unique label
\end{figure}

From Table \ref{tab-1}, we can see that the NCSMC g.s. energy quickly converges to the 
NCSM extrapolated value, unlike in the NCSM/RGM.  
%approach we are able
%to obtain the converged value (NCSM) of the binding energy of $^6$He g.s. 
This is  
due to the fact that %within this approach, the inclusion of 
the $^6$He NCSM eigenstate takes into
account the short range correlations and $^4$He core polarization that are missing when considering the cluster basis alone.
It is also important to note that, in contrast to the behavior offered by the NCSM, the
NCSMC recovers the correct extended asymptotic behavior of the wave function.  
%is obtained within the
%NCSMC in comparison of the limitations of the Harmonic Oscillator basis used in the NCSM. 
In Fig  
\ref{fig-1} such comparison is shown in a preliminary calculation at an $N_{max}=6$ model space.

%For figures use the syntax of figure~\ref{fig-1}

\begin{figure}
\centering
\includegraphics[width=11cm,clip]{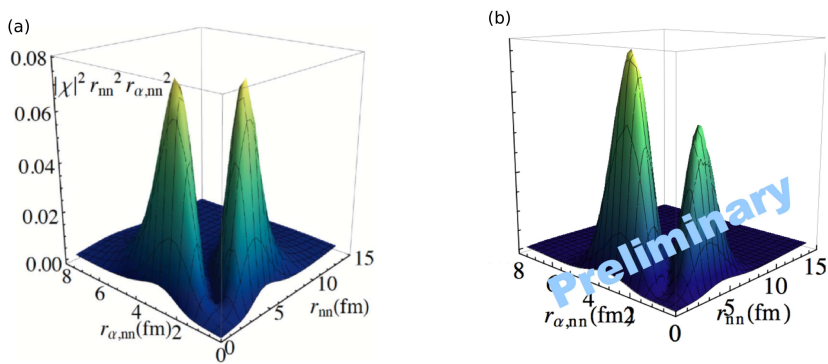}
\caption{Probability distribution of the $^6$He g.s. wave function in terms
of the relative distance between the neutrons ($r_{nn}$) and the distance 
between the center of mass of the neutrons and the $^4$He ($r_{\alpha,nn}$).
 The di-neutron and cigar configurations appear to have the same probability within
the NCSM/RGM (a), while the di-neutron probability is enhanced when using the NCSMC (b).   
%In panel (a), the NCSM/RGM calculation shows similar probability for both configurations while in panel
%(b), the NCSMC results enhance the strenght of the di-neutron with respect to the cigar configuration. 
%The characteristic  
%asymmetry (the di-neutron configuration more likely than the cigar) only appers when including short
 }
\label{fig-2}       % Give a unique label
\end{figure}

Finally, we can also compare the probability densities from the $^6$He g.s. obtained with the NCSM/RGM and
the NCSMC. In Fig. \ref{fig-2}, such comparison is shown and it is interesting to find that while
%the peaks of probability for 
the two main configurations (di-neutron and cigar) appear to have the same probability within  
the NCSM/RGM, the di-neutron probability is enhanced when using the NCSMC. This  
 asymmetry in the strength of the probability peaks is known to be a characteristic of $^6$He and
these results show that it is a consequence of the  
short range correlations.
 
\section{Conclusions}

The NCSMC uses an ansatz wave function that includes both an expansion in a continuum three-cluster basis 
and in a discrete basis of NCSM eigenstates. This provides a foundation that is capable of describing 
both short and long range characteristics of three-cluster systems. In the case of the $^6$He g.s., 
we  could see that this approach provides both the correct
binding energy and extended asymptotic behavior unlike the NCSM that does provide the correct binding  
energy, but not the correct asymptotics, or the NCSM/RGM that does the opposite.
Calculations in larger model spaces for both g.s. and continuum states  
of $^6$He are underway.

\begin{acknowledgement}
Prepared in part by LLNL under Contract DE-AC52-07NA27344. This material is based upon work supported by the U.S. Department of Energy, Office of Science, Office of Nuclear Physics, under Work Proposal Number SCW1158. Support from the
NSERC Grant No. 401945-2011 is acknowledged.
TRIUMF receives funding via a contribution through the
Canadian National Research Council.

\end{acknowledgement}

%
% BibTeX or Biber users please use (the style is already called in the class, ensure that the "woc.bst" style is in your local directory)
\bibliography{romero}
\end{document}